\def\year{2021}\relax
\documentclass[letterpaper]{article} 
\usepackage{aaai21}  
\usepackage{times}  
\usepackage{helvet} 
\usepackage{courier}  
\usepackage[hyphens]{url}  
\usepackage{graphicx} 
\usepackage{natbib}  
\usepackage{caption} 
\usepackage{amsmath}
\usepackage{algorithm}
\usepackage[noend]{algpseudocode}
\usepackage{makecell}
\usepackage{subfig}
\usepackage{multirow}
\usepackage{booktabs}
\usepackage{xcolor}
\usepackage{enumerate}
\usepackage{balance}
\usepackage{upgreek}

\title{How-to Present News on Social Media:\\ A Causal Analysis of Editing News Headlines for Boosting User Engagement}

\author{Kunwoo Park\textsuperscript{\rm 1}, Haewoon Kwak\textsuperscript{\rm 2}, Jisun An\textsuperscript{\rm 2}, Sanjay Chawla\textsuperscript{\rm 3}}
\affiliations{
\\
    \textsuperscript{\rm 1}School of AI Convergence, Soongsil University, South Korea\\
    \textsuperscript{\rm 2}Singapore Management University, Singapore\\
    \textsuperscript{\rm 3}Qatar Computing Research Institute, Qatar\\
    kunwoo.park@ssu.ac.kr, haewoon@acm.org, jisun.an@acm.org, schawla@hbku.edu.qa
}

\begin{document}
\maketitle

\begin{abstract}
To reach a broader audience and optimize traffic toward news articles, media outlets commonly run social media accounts and share their content with a short text summary. Despite its importance of writing a compelling message in sharing articles, the research community does not own a sufficient understanding of what kinds of editing strategies effectively promote audience engagement. In this study, we aim to fill the gap by analyzing media outlets' current practices using a data-driven approach. We first build a parallel corpus of original news articles and their corresponding tweets that eight media outlets shared. Then, we explore how those media edited tweets against original headlines and the effects of such changes. To estimate the effects of editing news headlines for social media sharing in audience engagement, we present a systematic analysis that incorporates a causal inference technique with deep learning; using propensity score matching, it allows for estimating potential (dis-)advantages of an editing style compared to counterfactual cases where a similar news article is shared with a different style. According to the analyses of various editing styles, we report common and differing effects of the styles across the outlets. To understand the effects of various editing styles, media outlets could apply our easy-to-use tool by themselves.
\end{abstract}

\section{Introduction}

People prefer to read their news online rather than newspapers these days~\cite{mitchell2018americans}.
This paradigm shift has brought both good and bad influences on the news industry. 
The bad is that the competition among news organizations has become intense. 
Since the distribution cost of news content is far less expensive than it used to be in the pre-digital news era, many online news media have newly appeared, and the amount of news stories published in a day has been soaring~\cite{atlantic2016}.
The good, on the other hand, is that it enables media to get direct feedback from their audience; it further makes it easier to \textit{quantitatively} measure the level of engagement on each news article. 
News organizations are increasingly adopting data-driven methods to understand their audience preferences, decide the coverage, predict article shelf-life, or recommend next articles to read~\cite{castillo2014characterizing,an2017multidimensional,kuiken2017effective,aldous2019view}. Data-driven methods have also increased the understanding of effective news headlines that boost traffic~\cite{kuiken2017effective,Hagar2019} while some headlines could undermine the credibility of news organizations in return for increased traffic~\cite{chen2015misleading}. 

\begin{figure}[t]
\centering
\includegraphics[width=.9\columnwidth]{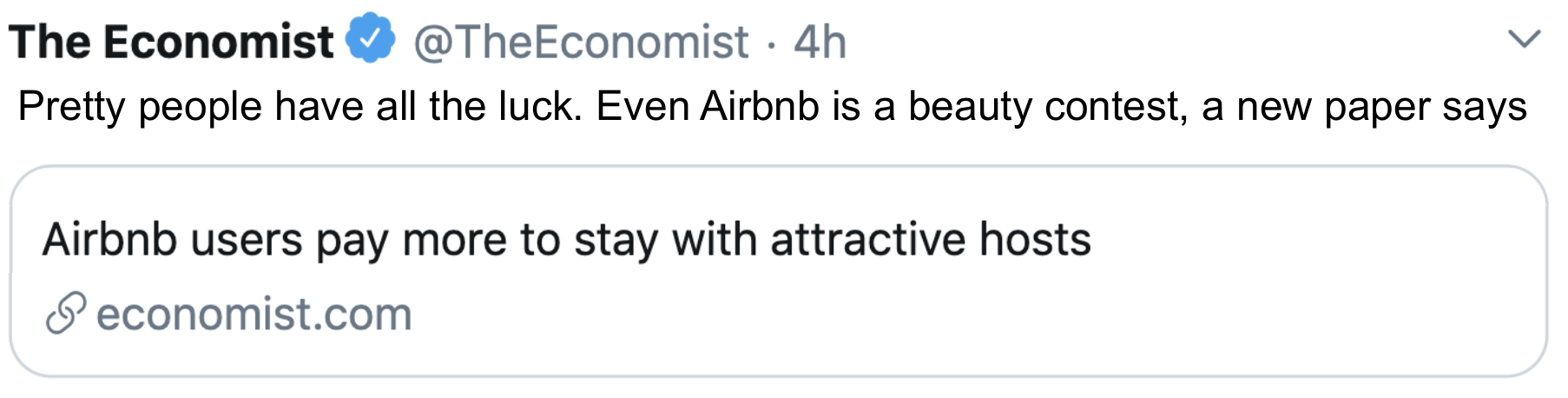}
\caption{An example of news article shared by a media outlet on Twitter.}
\label{fig:example}
\end{figure}

Sharing news articles on social media is a well-known strategy for boosting traffic to online news outlets. As shown in Figure~\ref{fig:example}, news outlets run their official accounts (we call \textit{media account} for the rest of this paper) and share a link to a news article with a short text. 
There are various ways for writing the short message. 
One way is to mirror an original news headline without any modification because the news headline is a concise summary of what the news article is about~\cite{van2013news}. The other way is to edit the news headline considering more informal characteristics of social media~\cite{welbers2019presenting}, such as adding clickbait-style phrases like ``Even Airbnb is a beauty contest'' in the figure.
Social media managers at the newsroom face such a challenging task every day about how to write a more effective message for news sharing~\cite{aldous2019challenges}.
In spite of its importance, they depend on their experience and make educated guesses to maximize audience engagement. As a result, they sometimes fail. Also, the research community is aware of different practices of using social media across media outlets~\cite{russell2019twitter,welbers2019presenting} but does not own a sufficient level of understanding on which strategies lead to increased user engagement.

In this work, we aim to fill this gap by analyzing news articles shared on Twitter by eight news media outlets, which have diverse publication channels and political leaning. In particular, we tackle the following research questions to deepen our understanding of editing practices of media accounts and their effects:
\begin{enumerate}[RQ\(1\).]
    \item \textit{How do news media edit news headlines when sharing news articles on social media?}
    \item \textit{Which kind of editing style leads to more audience engagement on social media?}
\end{enumerate}
We characterize how media accounts write a tweet against its original news headline and evaluate its effectiveness on the amount of user engagement, such as the number of retweets or likes, by using a systematic framework that incorporates propensity score analysis with deep learning. 

The main contributions of this paper are as follows:
\begin{enumerate}
    \item We build a parallel text corpus of news articles and social media posts (tweets in this work) written by eight hybrid and online-only media accounts and make it publicly available to a research community. From the dataset, we characterize patterns on how media outlets edit tweet messages when sharing news articles on Twitter.
    \item To estimate the effects of editing news headlines on user engagement, we utilize a systematic framework that employs a deep learning-based model for propensity score analysis: it compares the level of user engagements for a style with counterfactual cases where similar news articles are shared with a different editing style. This framework can be applied to any paired dataset of news articles and  social media messages, offering practical contributions to news media outlets for evaluating how effective their strategy of publishing social media messages is.
    \item Using the analysis framework on the dataset of the eight news outlets, we test which kind of editing strategy is effective in audience engagement. 
    While we observe that sharing a clickbait-style tweet achieves a larger amount of user engagement compared to its estimated counterfactual cases for half of the media outlets in this study, the opposite effect---a clickbait tweet decreases user engagement---is also found for other media.
\end{enumerate}

\section{Related Works}

\subsection{News Media in the Era of Social Media}

There has been a line of research on how news organizations use social media in terms of content and interaction. News organizations use Twitter as a promotional tool and write a tweet of news headlines with a corresponding link~\cite{armstrong2010now,holcomb2011mainstream}. Another study pointed out that news media employ their accounts as a mere news dissemination tool without much interaction with audience~\cite{malik2016macroscopic}.
However, the current practices on using social media vary across the news media and countries~\cite{russell2019twitter,welbers2019presenting}. 

The emergence of social media also brings changes in news writing~\cite{dick2011search,tandoc2014journalism}, particularly in news headlines.
In traditional newspapers, news headlines are expected to provide a clear understanding of what the news article is about~\cite{van2013news} for helping those who read a newspaper while scanning headlines. Hence, headlines have functioned as a summary of the key points of the full article~\cite{bell1991language,nir1993discourse}.
As social media become popular~\cite{kwak2010twitter,hermida2012share}, headlines are also required to attract readers' attention to increase traffic to their websites~\cite{chen2015news}.
Accordingly, editors and journalists have adjusted the way they write headlines~\cite{dick2011search}. The characteristics of headlines in online news have been studied across the platforms, styles, sentiments, and news media~\cite{kuiken2017effective,dos2015breaking,scacco2019curiosity,piotrkowicz2017headlines}.

\subsection{News Popularity and User Engagement}
\label{sec:related_work:news_popularity}

A significant amount of work has attempted to predict the popularity of news articles on web environments by modeling content features of news articles and user reactions on news websites and social media.
Various studies have concluded that early user reactions on social media have a strong predictive power for the long-term popularity of news articles~\cite{lerman2010using,castillo2014characterizing,keneshloo2016predicting}. 
Bandari et al. (\citeyear{bandari2012pulse}) tackled a more challenging problem in forecasting the popularity (mainly view counts) of news articles even before its publication, which is known as `cold start' prediction.
However, applying the popularity prediction models relying only on news content was not successful for the cold-start prediction in practice~\cite{arapakis2014feasibility}. The importance of delivering fresh news earlier than competitors to attract readers is reported~\cite{rajapaksha2019scrutinizing}. 
In addition to views, various dimensions of audience engagements have been studied. Tenenboim and Cohen (\citeyear{tenenboim2015prompts}) compared the most-clicked items with the most-commented items and found that 40-59\% of the items are different. Aldous et al. (\citeyear{aldous2019view}) reported varying topical effects on user engagement across engagement types, such as views, likes, and comments.

Kuiken et al. (\citeyear{kuiken2017effective}) investigated the impact of editing a news headline with clickbait on view counts, which is a specific type of news headlines designed for attracting users' attention by using a catchy text~\cite{chen2015misleading} or referring content that is not exposed in a headline~\cite{blom2015click}. 
From one Dutch news aggregator, Blendle, they examined 1,828 pairs of the original news headline and the rewritten title by Blendle editors. They found that rewriting a headline with clickbait is likely to increase the number of views. 

Another line of research examined the role of posting time for the popularity of news articles. Using regression analyses for predicting view counts of the Washington Post articles, Keneshloo et al. (\citeyear{keneshloo2016predicting}) showed that the posting time was not an important factor for audience engagement. Another study investigated social media messages shared by Twitter accounts of 200 Irish journalists~\cite{orellana2016spreading} and suggests that there is no best time of the day for engagements; they only found out a slight increase in audience engagement after 5 pm.

In the subsequent sections, we will first investigate how media accounts write tweets for sharing news articles on social media. Then, to estimate the effects of editing styles (e.g., mirroring news headlines or adding clickbait phrases), we will apply a systematic framework that controls for the effects of confounding variables on engagement. Following the literature on news popularity and audience engagement, we decide to control for the effects of news content as a major confounding variable in the following analyses.

\section{Data Collection}

To answer our research questions, we first build a parallel text corpus of news articles and social media posts. For covering diverse posting styles, we consider two types of news media in terms of channels for publishing news: hybrid news media and online-only news media. Hybrid news media (e.g., CNN) are the news outlets with both conventional mass media channels, such as newspapers and television and online channels. By contrast, online-only news media (e.g., HuffPost) are emerging media that publish content through online channels only.

\begin{table}[t]
\frenchspacing \footnotesize
\centering
 \begin{tabular}{ccrr}
    \toprule
    Type & Media & Followers & Tweets \\
    \midrule
    \multirow{4}{*}{Hybrid} & The New York Times & 43.7M & 143,011 \\
    &The Economist & 23.8M &  30,200 \\
    &CNN & 42.2M & 50,841 \\
    &Fox News & 18.5M  & 34,245 \\
    \midrule
    \multirow{4}{*}{Online-only} & HuffPost & 11.4M & 23,712 \\
    &ClickHole & 487K & 5,535 \\
    &Upworthy & 516K & 168 \\
    &BuzzFeed & 6.56M & 18,862 \\    
    \bottomrule
\end{tabular}
\caption{Descriptive data statistics}
 \label{tbl:data_summary}
 \end{table}

For hybrid news media, we collect a list of reliable news media and their political leaning from Media Bias/Fact Check~\cite{mediabiasfactcheck}, which is widely used in large-scale news media analysis. 
We also manually compile their social media accounts and their number of followers on social media. 
We then choose four popular news media to include different political leanings in our dataset: The New York Times (@nytimes, left-center), The Economist (@TheEconomist, least-biased), CNN (@CNN, left), and Fox News (@FoxNews, right). The popularity is measured based on the number of followers on Twitter.
For online-only news media, we choose four news media: HuffPost (@HuffPost), ClickHole (@ClickHole), Upworthy (@Upworthy), and BuzzFeed (@BuzzFeed) based on previous literature~\cite{chakraborty2016stop} and their popularity.
For these eight media outlets, our data collection pipeline consists of four steps:

\begin{enumerate}[(1)]
    \item We collect tweets written by each media account. Using \textit{twint}\footnote{https://github.com/twintproject/twint}, a third party library for Twitter data collection, we collect all available tweets but not mentions nor retweets. We also exclude tweets that contain an URL only without any text.
    
    \item We extract an embedded URL from each tweet. As it is typically shortened (e.g.,  http://nyti.ms/2hKFRvl) and sometimes shortened multiple times, we expand it until it reaches the final destination. If the expanded URL points to other sites, such as YouTube, we exclude it.
    
    \item We retrieve the HTML document of expanded URLs pointing to news articles. Our crawler sends requests with generous intervals.
    
    \item As the last step, we extract a pair of headline and body text from each HTML file we collected. 
\end{enumerate}

Table~\ref{tbl:data_summary} is the summary statistics of our dataset used in this work. Our dataset consists of the pairs of news articles and their tweets that were published in 2018. For the New York Times, we utilize a publicly available corpus~\cite{fakenewscorpus2017} for Step (3) and match news articles with their tweets in 2018. The completeness of this corpus has been reported~\cite{kwak2020systematic}.
We also note that Upworthy actively tweeted only in the last two months of 2018. Due to the copyright issues, we only share news headlines accompanied with its corresponding tweet ids at the following repository\footnote{https://github.com/bywords/NTPairs}. One can easily retrieve our paired dataset by hydrating the tweets using the official Twitter API or third party libraries with the provided tweet ids.

\section{How News Media Edited Tweets}

To understand how media accounts edit tweet messages when sharing news articles on social media (RQ1), we characterize media accounts from the perspectives of headline mirroring, content change in lexicons and semantics, and clickbaitness of headlines and tweets.

\subsection{How Often Do Media Accounts Mirror Headlines?}

\begin{table}[t]
\centering
\subfloat[Hybrid media]{
\begin{tabular}{cccc}
\toprule
\makecell{NYTimes} & \makecell{TheEconomist} & \makecell{CNN} & \makecell{FoxNews}\\\midrule
0.12310 & 0.12417 & 0.07543 & 0.39731\\\bottomrule
\end{tabular}
}
\\
\subfloat[Online-only media]{
\begin{tabular}{rrrr}
\toprule
\makecell{Huffpost} & \makecell{ClickHole} & \makecell{Upworthy} & \makecell{BuzzFeed}\\\midrule
0.00017 & 0.80976 & 0.71429 & 0.29350\\\bottomrule
\end{tabular}
}
\caption{Fraction of the tweets with the mirroring headlines (Twitter handles are presented)}
\label{tbl:frac_mirroring}
\end{table}

Considering that the mainstream news media outlet publishes about 150 to 500 news stories per day~\cite{atlantic2016}, it may be challenging for news outlets to write new social media text for all their news stories. 
Thus, using news headlines, which are already a good and concise summary for news articles, without any modification (so-called mirroring) might be a reasonable choice for news media.
We first examine how often media accounts mirror a news headline and edit the headline to better appeal to social media users. Table~\ref{tbl:frac_mirroring} presents the proportion of the mirrored headlines across the media outlets. Of the 8 news media, Huffpost is the most active in editing headlines for social media; only 0.017\% of the tweets contain the original headlines. By contrast, ClickHole mirrors the original headlines in 80.976\% of their tweets. 
Then, when a change happens, how much content of the headline is preserved in the tweet, and how its degree differs across the media outlets?

\subsection{How Much Is the Content Preserved?} 

\begin{figure}[t]
\centering
\includegraphics[width=.9\columnwidth]{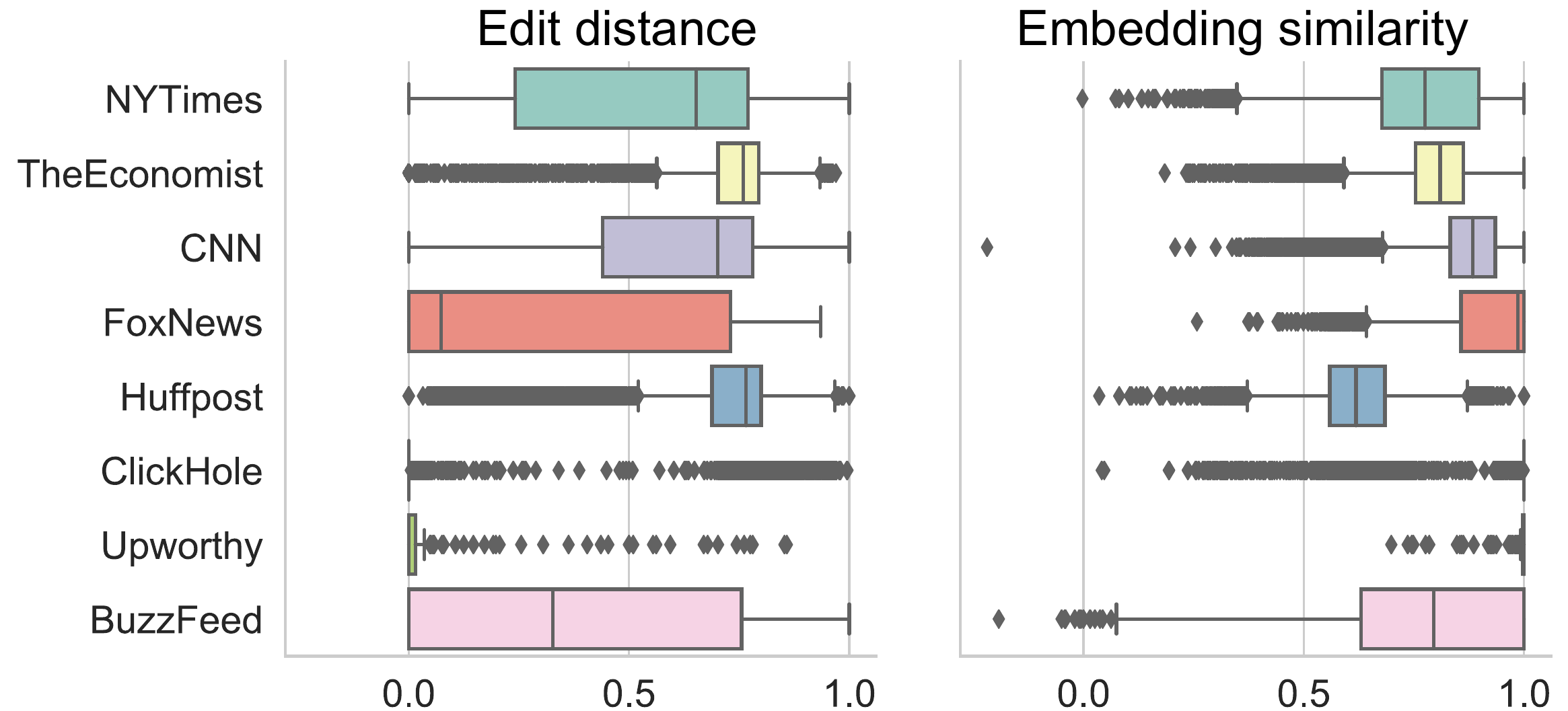}
\caption{Edit distance and embedding similarity between news headlines and tweets}
\label{fig:sen_emb_similarity_bleu_w_outliers}
\end{figure}

To examine how media accounts preserve original headlines when editing tweet messages, we use two measures that quantify the similarity between a news headline and the corresponding tweet text: Levenshtein distance (\textit{edit distance}) and Cosine similarity over an embedding space (\textit{embedding similarity}). 
First, edit distance is utilized to quantify how many edits (deletion, insertion, and substitution) are required to transform a news headline into a tweet text. We normalize edit distance by the longer length of the two texts, ranging from 0 (identical) to 1 (no character overlap). 
Second, to know whether how much semantics are preserved, we measure embedding similarity by utilizing a pre-trained fastText word embedding~\cite{fasttext_english}. 
We map a headline and its corresponding tweet into 300d vectors using the embedding and measure the cosine similarity between the two vectors, ranging from -1.0 (dissimilar) to 1.0 (identical). Contrary to edit distance, a higher score indicates that the two texts are more similar to one another.

Figure~\ref{fig:sen_emb_similarity_bleu_w_outliers} shows the degree of content preservation of the eight news outlets, which is measured by edit distance and embedding similarity. Not surprisingly, most media accounts tend to make a small amount of change for posting tweets against its original news headline, which are represented as a high value of embedding similarity and a low value of edit distance. 
However, some outlets exhibit distinct patterns; for example, in HuffPost, the median value of embedding similarity is only 0.619, which is significantly lower than the overall median value of 0.835. We further investigate the media-level difference by employing a Mann-Whitney's U test between each pair of the eight outlets on edit distance and embedding similarity, respectively. 
All of the pairwise relationships show statistically significant differences (\textit{p}$<$.0001) except one pair of ClickHole and Upworthy (\textit{p}$=$0.554).
Taken together, the above observations suggest that the media outlets have their own writing styles for news sharing on Twitter.

By examining edit distance and embedding similarity simultaneously, we can reveal a more detailed picture of how much the content of a news headline is preserved in its corresponding tweet. 
For example, if edit distance is low but embedding similarity is high, the tweet should be almost identical to the headline. 
By contrast, if both edit distance and embedding similarity are high, the tweet may preserve the meaning but is written very differently against the headline, which corresponds to a paraphrase of the headline.

\begin{figure}[t]
\centering
\subfloat[New York Times]{\includegraphics[width=0.24\textwidth]{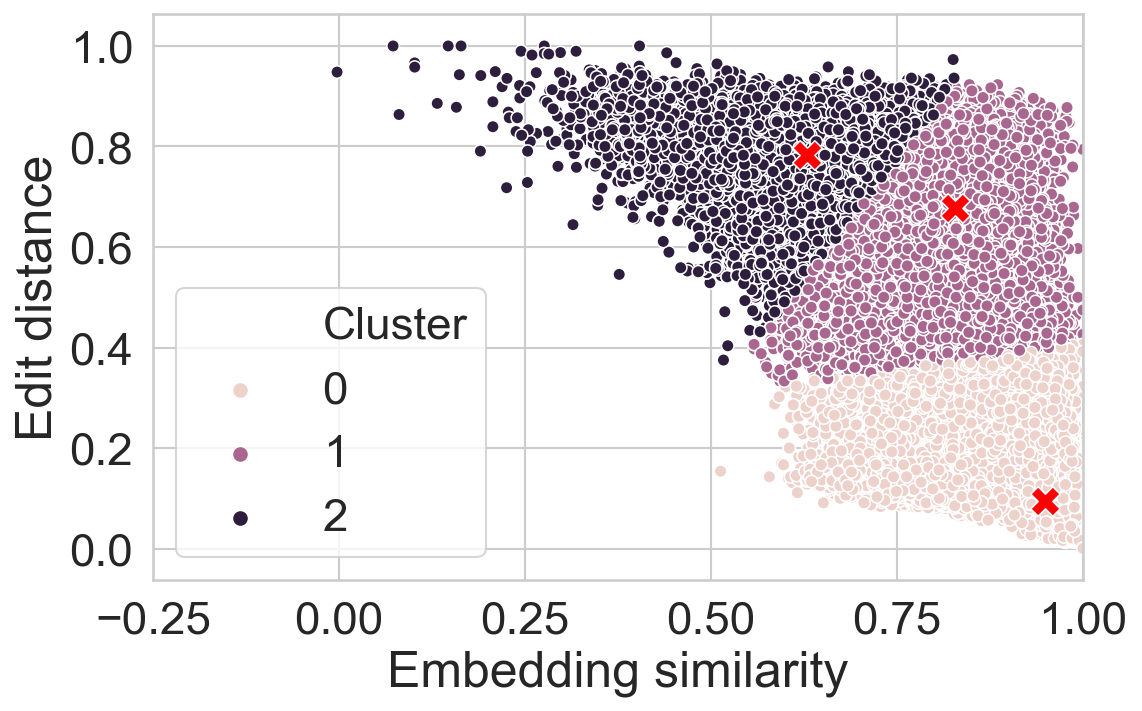}} 
\subfloat[CNN]{\includegraphics[width=0.24\textwidth]{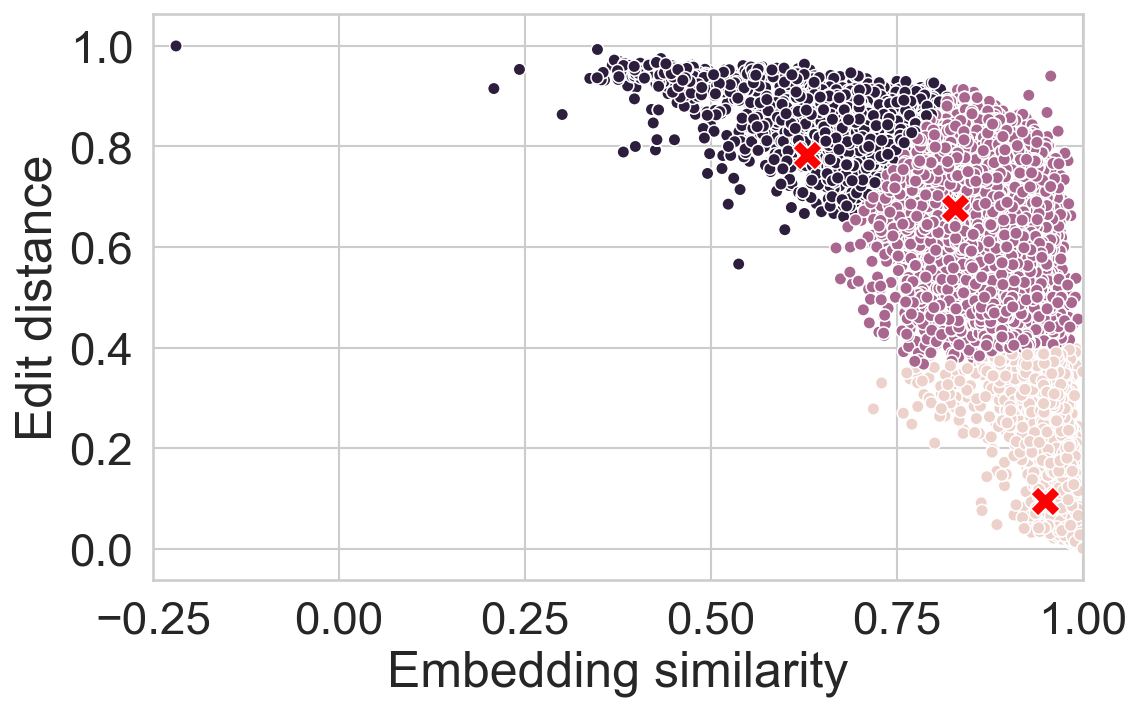}}\\
\subfloat[HuffPost]{\includegraphics[width=0.24\textwidth]{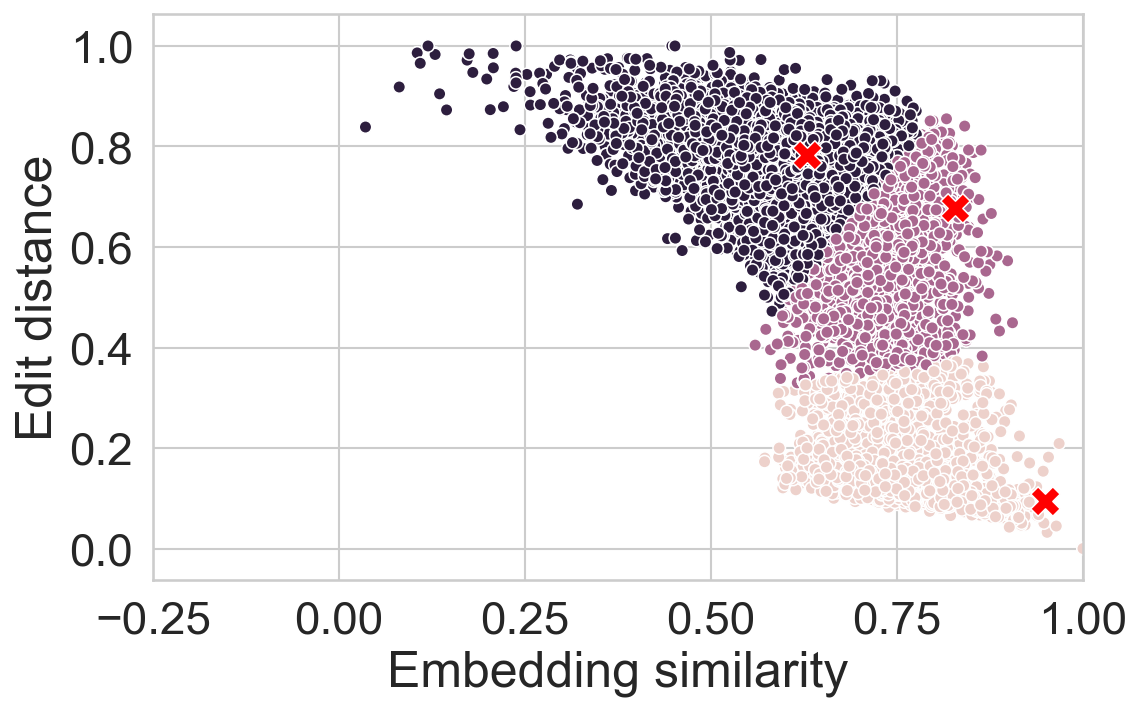}} 
\subfloat[BuzzFeed]{\includegraphics[width=0.24\textwidth]{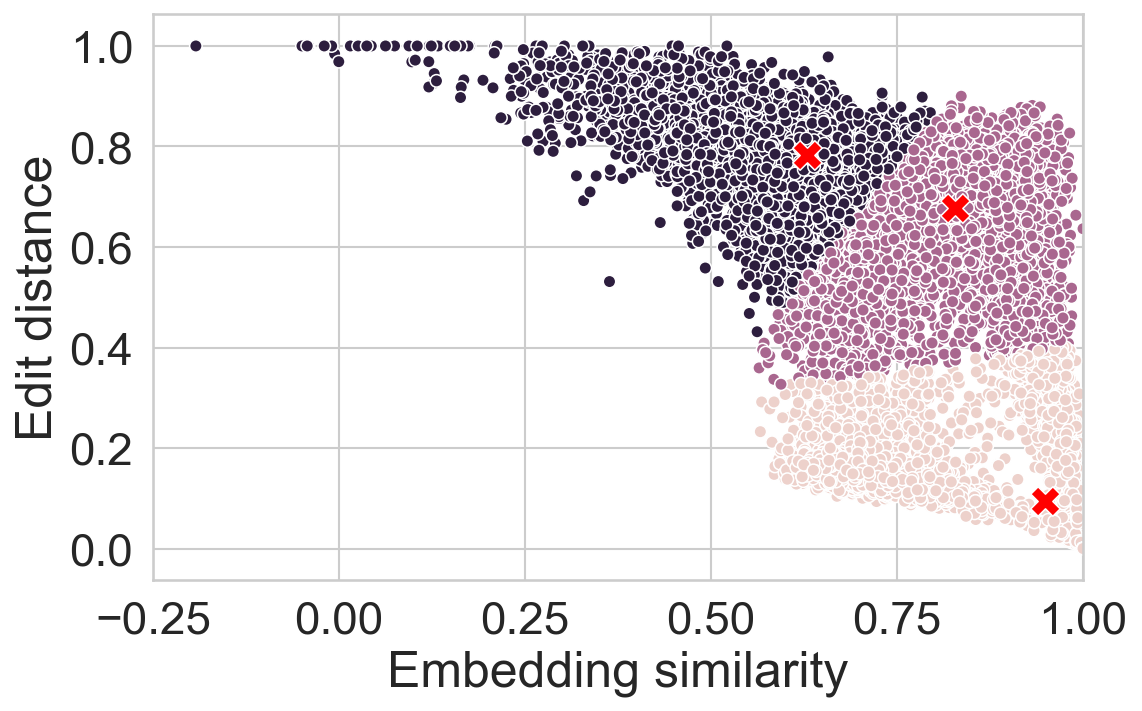}}\\
\caption{Identified clusters of (news headline, tweet) pairs by edit distance and embedding similarity}
\vspace{-2mm}
\label{fig:cluster}
\end{figure}

To figure out how the two measures interact and identify common or differing patterns across the eight media outlets, we draw the scatter plots of edit distance and embedding similarity in Figure~\ref{fig:cluster}. 
Each dot indicates a change between a news headline and its corresponding tweet, which represents embedding similarity along the $x$-axis and edit distance along $y$-axis.
To aggregate similar patterns into a handful number of groups, we apply the K-means++ clustering algorithm, which improves the standard K-means by assigning initial centroids based on the underlying data distribution, to the whole data. We determine the optimal number of clusters ($k$=3) by the elbow method. In the figure, each dot's color indicates the cluster index, and the red $X$ marks are the centroids of each cluster. Due to the lack of space, we present the results of the two outlets each from hybrid and online-only media by data size. Here, we do not argue that the identified clusters represent general editing styles; instead, through the proxies, this approach enables us to systemically understand how news media outlets edit tweets in terms of lexicons and semantics. Further studies could characterize editing patterns through a combination of quantitative analysis and qualitative investigation on multiple datasets.

\begin{table}[t]
\centering
\begin{tabular}{c|rrr}
\multirow{1}{*}{Media} & \makecell{Cluster 0} & \makecell{Cluster 1} & \makecell{Cluster 2}\\\hline
NYTimes & 0.3173 & 0.3246 & \textbf{0.3581}\\
TheEconomist & 0.1332 & \textbf{0.6095} & 0.2572\\
CNN & 0.2236 & \textbf{0.6864} & 0.09 \\
FoxNews & \textbf{0.6478} & 0.2866 & 0.0656 \\\hline
Huffpost & 0.0987 & 0.1064 & \textbf{0.7949} \\
ClickHole & \textbf{0.8499} & 0.0094 & 0.1407 \\
Upworthy & \textbf{0.8929} & 0.0952 & 0.0119 \\
BuzzFeed & \textbf{0.5132} & 0.1787 & 0.3081 \\\hline
\end{tabular}
\caption{Fraction of the clusters determined by edit distance and embedding similarity between headline and tweet}
 \label{tbl:cluster_fraction}
 \end{table}
 
Table~\ref{tbl:cluster_fraction} demonstrates the fraction of headline-tweet pairs that belong to each cluster. Cluster $0$ represents the pairs of which a tweet is similar or identical to the news headline, having low edit distances and high embedding similarities. In Cluster $1$, many lexical changes are made, but the semantics of a tweet is still similar to the corresponding news headline, as represented by high edit distances and high embedding similarities. This pattern suggests that Cluster $1$ may indicate paraphrasing. Cluster $2$ demonstrates the highest edit distance and the lowest embedding similarity, suggesting that a tweet may be re-written with less similar semantics for sharing news articles on Twitter.

Here, we make observations on common editing patterns against the type of media. Cluster $0$ is the most frequent group for those online-only media except for Huffpost. Incorporated with the patterns in Table~\ref{tbl:frac_mirroring}, the observations show that the online-only media tend to share news headlines with a marginal amount of change. 
On the other hand, the hybrid media tend to rarely employ editing styles represented by Cluster $0$, except for FoxNews: Cluster $1$ is the most frequent for TheEconomist and CNN while NYTimes shows a balanced distribution over the clusters. This finding suggests that the hybrid media outlets actively rewrite messages for sharing news articles on social media, while each outlet may use distinct styles as represented by the varying cluster distribution.

\subsection{How Differently Do News Outlets Use Clickbait-Style Headlines and Tweets?}

As the news industry becomes competitive, news outlets have published articles with a specific headline style that leads to more clicks by stimulating psychological perspectives, which is known as clickbait~\cite{kilgo2016six,molek2013towards,stroud2017attention}. While media credibility might be undermined when news outlets exploit clickbait too often in their websites, presenting clickbait might be acceptable on social media where people write casual expressions more frequently (e.g.,  Figure~\ref{fig:example}). To investigate how clickbait usages vary across the media outlets, we utilize a deep learning classifier that infers the patterns in headline-tweet pairs.

Using a public dataset of clickbait and non-clickbait headlines that were manually annotated in \cite{chakraborty2016stop}, we first train an attention-based bidirectional recurrent neural network (RNN) classifier. The gated recurrent unit (GRU) is used as a basic unit, and an attention mechanism is, in turn, applied to the RNN hidden units. 
We train the network to minimize the cross-entropy loss using Adam optimizer with gradient clipping. On a separate test set of 90:10 split, the model achieves an F1-score of 0.994, which outperforms the baseline performance of 0.934 using SVM~\cite{chakraborty2016stop}. Using the classifier, we estimate the clickbait score of a tweet or a headline by using the sigmoid output between 0 and 1.

\begin{figure}[t]
\centering
\includegraphics[width=0.85\columnwidth]{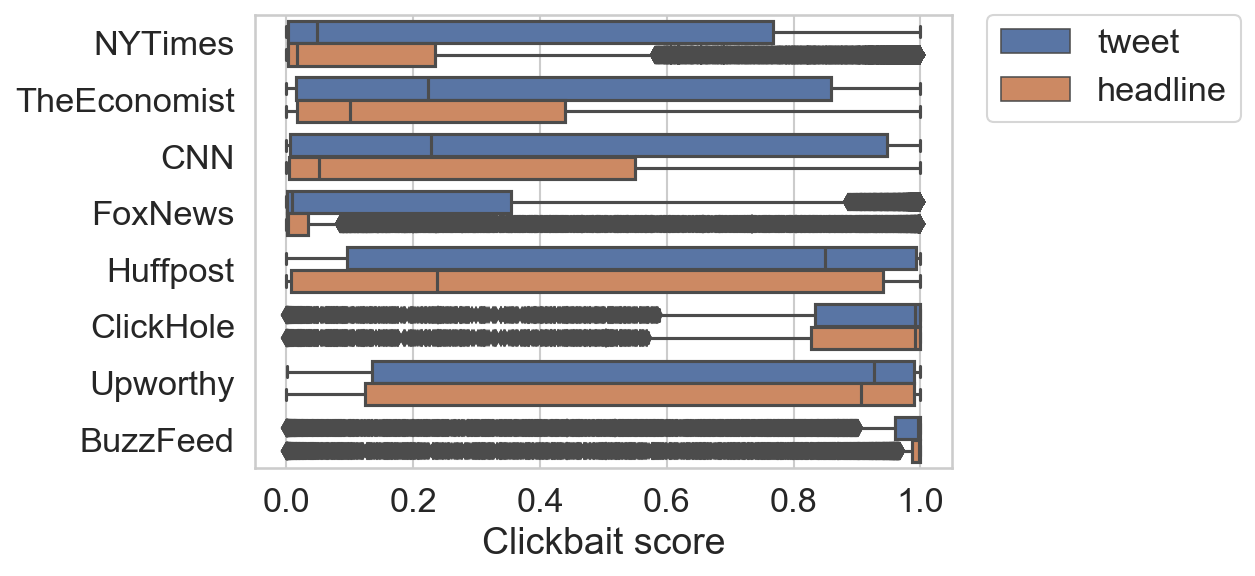}
\vspace{-2mm}
\caption{Clickbait scores of news headlines and tweets}
\vspace{-2mm}
\label{fig:clickbait_score}
\end{figure}

Figure~\ref{fig:clickbait_score} demonstrates the distribution of clickbait scores for news titles and tweets of the eight news media. Hybrid news media are less likely to use clickbait on news headlines. On the contrary, the clickbait scores in tweets are higher than those in headlines (\textit{p}$<$0.001 by t-test). This observation implies that the hybrid media tend to publish news articles with non-clickbait headlines but try to share them with more clickbait tweets to be adapted to the social platform. 
By contrast, the online-only media use clickbait actively, both for headlines and tweets. Among them, tweets of Huffpost and BuzzFeed have high clickbait scores in tweets compared to their headlines (\textit{p}$<$0.001 by t-test).  

The above results reveal the common and differing trends in the usage of clickbait across the media. In original news articles, hybrid media tend to use clickbait less frequently in headlines, yet online-only media employ actively. On the other hand, most of the media use more catchy expressions in sharing news on Twitter.

\begin{table}[t]
\centering
\subfloat[][Hybrid]
{
\begin{tabular}{c|cc}
Media & $P(NC|C)$ & $P(C|NC)$ \\\hline
NYTimes & 0.188 & 0.226 \\
TheEconomist & 0.381 & 0.333 \\
CNN & 0.222 & 0.301 \\
FoxNews & 0.2 & 0.156 \\\hline
Macro Average & 0.248 & 0.254 \\\hline
\end{tabular}
}
\\
\subfloat[][Online-only]
{
\begin{tabular}{c|cc}
Media & $P(NC|C)$ & $P(C|NC)$ \\\hline
Huffpost & 0.167 & 0.474 \\
ClickHole & 0.036 & 0.133 \\
Upworthy & 0.016 & 0.103 \\
BuzzFeed & 0.054 & 0.333 \\\hline
Macro Average & 0.068 & 0.261 \\\hline
\end{tabular}
}
\caption{P(Tweet$_{class}$ $\vert$Headline$_{class}$) for hybrid and online-only media  (C=Clickbait, NC=Non-clickbait)}
\label{tab:conditional_clickbait}
\end{table}

To better understand how each outlet exploits clickbait when sharing a news article on Twitter, we compute the probability of shifting the clickbaitness of news article when sharing it on social media: 
P(Tweet$_{class}$ $\vert$Headline$_{class}$)\footnote{Text$_{class}$ is clickbait when the clickbait score of the text $>$ 0.5.}. 
Table~\ref{tab:conditional_clickbait} reports the conditional probability of a tweet to be clickbait or non-clickbait given the headline is clickbait or non-clickbait. Here, we make common and different trends for the media group. Given a non-clickbait news headline, the probability of its tweet to be clickbait is similar across the hybrid and the online-only media (0.254 and 0.261, respectively). On the other hand, when the original news headline is clickbait, the hybrid and online-only media accounts shift the style with a huge difference. While the hybrid media flips the news headline's clickbaitness with a similar probability compared to the cases when non-clickbait news (0.248), the online-only media rarely do (0.068). This observation implies the online-only media prefer to use clickbait tweets in any case.

\section{Effects of Editing Styles for News Sharing on User Engagement}

In the previous section, we present that the eight news media outlets employ various strategies in editing tweets when sharing their news articles on Twitter.
While the simplest tactic of the media accounts is to mirror the original headline of a news article on social media, news media also make a significant amount of edits with changes of content and text styles. Then, which strategies would be more effective for user engagement on social media? How should media accounts write a tweet message (RQ2)?

\begin{figure}[t]
\centering
\includegraphics[width=0.85\columnwidth]{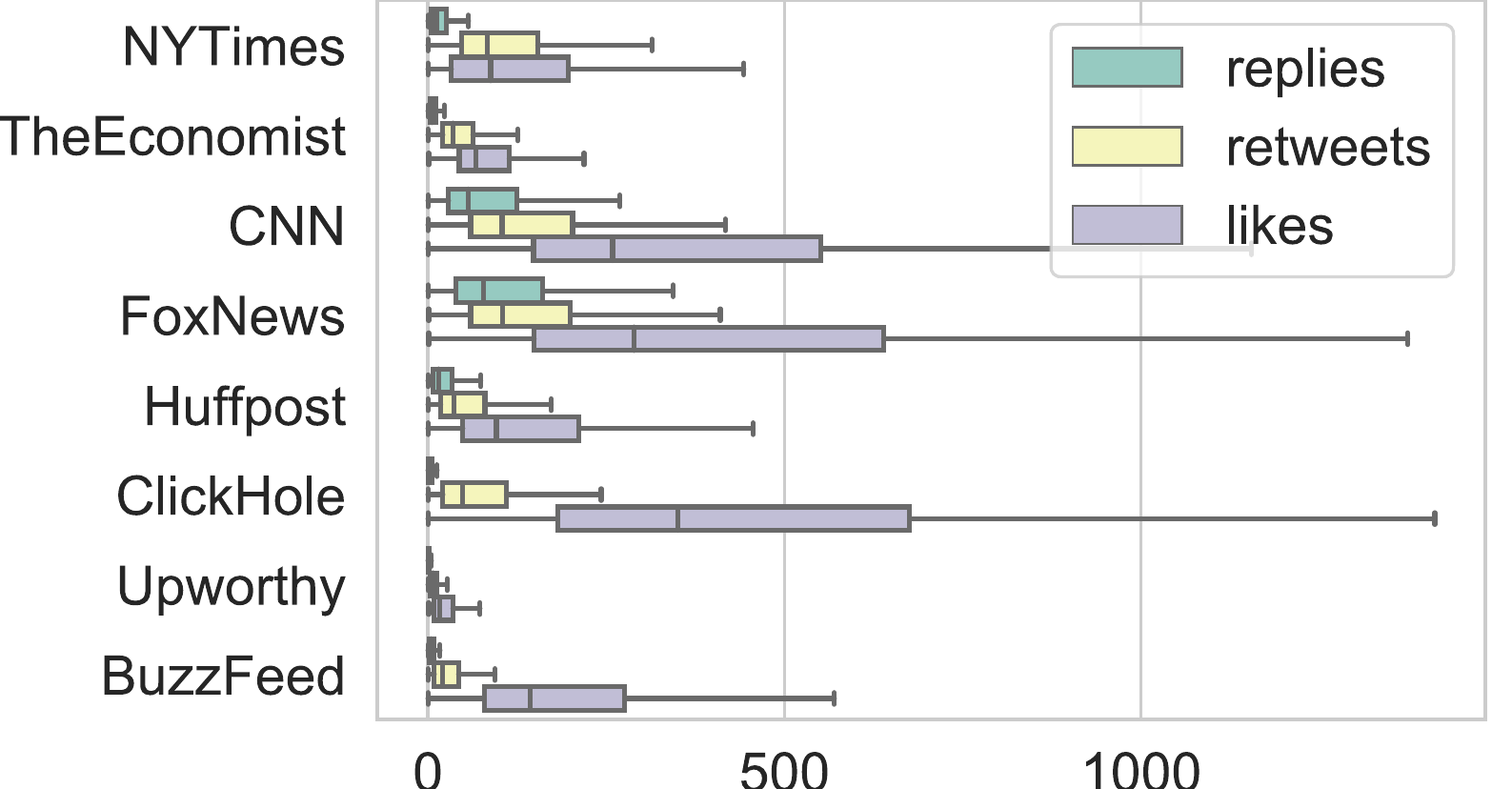}
\caption{User reactions on the tweets published by the media accounts (w/o outliers)}
\label{fig:reaction}
\end{figure}

We use the number of replies, retweets, and likes as a proxy of user engagement. Figure~\ref{fig:reaction} shows the distribution of the three metrics. Among the eight media outlets, FoxNews garnered the highest amount of user engagement across the three variables. ClickHole harvested the equivalent amount of likes to that of FoxNews but got lower number of retweets and replies. This observation suggests that each of the three measures reflects a different aspect of user engagement on Twitter, and thus the effects of editing tweets should be analyzed separately for each of the engagement metrics and the news outlets.

\subsection{Analysis Framework}

We utilize a systemic framework that incorporates propensity score analysis~\cite{rosenbaum1983central} with deep learning. The propensity analysis framework is widely used for estimating a causal effect of having a treatment condition from an observational dataset. To test whether a certain causal relationship exists from a treatment variable to an outcome variable, researchers generally conduct a controlled trial on human or animal subjects, for example, the effects of taking a pill on reducing the headache symptom. Since the casual relationship can be confounded by certain variables called covariates, such as gender and age, researchers randomly assign subjects into one of treatment group (taking a real pill) and control group (taking a placebo).

In observational studies where data is given, however, researchers cannot control the process of data generation; therefore, observing correlations between a treatment variable and an outcome variable can be confounded by covariates. In this study, for example, we aim at measuring the effects of a certain editing style for news sharing on audience engagement on Twitter; however, merely observing how the two variables are associated can be confounded by other factors such as news topics, which might affect the probability of that news media employ the editing style (Covariates$\rightarrow$Treatment) as well as the expected amount of engagement independent of editing styles (Covariates$\rightarrow$Outcome).

Propensity score matching (PSM) has been widely applied to observational studies on social media to address the issue~\cite{de2017language,olteanu2017distilling,park2020trust}. PSM first models a probability of having a treatment condition from given covariates (i.e., $P($Treatment$|$Covariate$)$). Next, PSM `matches' the instances of the corresponding control group to each treatment unit that have a propensity score similar to that of the treatment unit. This process approximates a randomized controlled trial in which the analysis units are randomly assigned into either treatment or control group, and thus, the risks of confounding effects due to covariates can be reduced. For more details of PSM, please refer to \citet{guo2014propensity}.

\subsubsection{Modeling Propensity Scores}

s discussed in related studies~\cite{tenenboim2015prompts,mummolo2016news}, the probability of selecting news items gets increased when a news article covers the topics of a reader's interest, and so does the likelihood of reacting to news shared on social media. Therefore, we aim at reducing the confounding effects of topics on audience engagement by modeling a deep learning-based propensity model that takes as input the body text of news articles. While social media engagements are also subject to who the posters are, we do not include it as one of the covariates because the analysis framework is applied to each news outlet separately; that is, the poster effects are controlled by the analysis design.

To model the propensity score, we employ deep learning techniques that have shown state-of-the-art performance in text classification tasks in recent studies. In particular, we first transform a sequence of words in body text into a 300-dimensional vector by averaging word vectors that were pre-trained using fastText~\cite{joulin2016bag} on a news dataset~\cite{fasttext_english}. The sentence vectors are fed into the three-layer fully-connected neural networks. We use the ReLU non-linearity, and the L2 regularization is applied to the last hidden layer ($\lambda$=0.001). We train the whole network by minimizing the cross-entropy loss of the treatment label and the predicted value.

\subsubsection{Matching}

The next step is to match each treatment unit to the control units based on the propensity score. To put it differently, we prune instances that are too different from treatment groups in terms of the propensity score. We apply the $k$-nearest neighbor algorithm ($k$=5) to each treatment unit. 
After the matching process is completed, the general PSM framework requires to check balances between a treatment group and its matched controls by the standardized mean difference of each covariate~\cite{guo2014propensity}. If the two groups are not balanced, we cannot proceed the rest step since they cannot satisfy the conditional independence assumption, which is required to estimate a causal effect. In our experiments of which the text feature is represented by a 300-d latent vector, we alternatively use the cosine similarity between the embedding vectors of the treatment and control units, which is widely used to measure the similarity between two documents in the NLP community~\cite{manning1999foundations}.

We formalize the condition of the successful matching as follows:
\begin{equation}
   \frac{1}{|T|}\sum_{t}^{T} \sum_{m}^{M_{t}} \frac{Similarity(t, m)}{k} \geq max(\mu+\alpha\times\sigma, \uptau)
\label{eq:condition}
\end{equation}
\noindent , where $T$ is a set of treatment units and $M_{t}$ is a set of control units matched to treatment unit $t$. $\alpha$ is a hyperparameter that controls the sensitivity of deciding whether a matching is successful, and $k=|M_t|$ is the number of neighbors for a treatment unit, which is set as a hyperparameter of the nearest neighbor algorithm. $\mu$ and $\sigma$ are the mean and standard deviation of embedding similarity between all the pairs of documents from the original dataset before matching. $\uptau$ is a thresholding value that copes with the distribution where similarities are on average low.
In the following experiments, we set $\alpha$ to be 1.5, which lets $\mu + \alpha\times\sigma$ corresponds to the 86th percentile of the similarity value, and $\uptau$ to be 0.8.

\subsubsection{Estimating Treatment Effects}

For the treatment groups with successfully matched instances, we estimate the treatment effect on user engagement. The Estimated Average Treatment Effect (EATE) on an outcome variable is measured as follows:
{
\begin{equation}
EATE=\sum_{t}^{T} \sum_{m}^{M_{t}} \Big( \frac{y_{t}-y_{m}}{k} \Big) /{N_{T}}
\label{eq:eate}
\end{equation}
}
\noindent, where $y_{t}$ and $y_{m}$ are the outcomes measured for $t$ and $m$, respectively. $N_{T}$ is the number of treatment units, and the meaning of other symbols is the same as those in Equation (\ref{eq:condition}). EATE quantifies the potential (dis-)advantage of user engagement by sharing a news article with a certain style (treatment) compared to another (control).

\subsubsection{Robustness Check Using Cross-Validation}

As discussed in \citet{kiciman2019causal}, it is crucial to conduct a sensitivity analysis for a successful propensity score matching because the matching process can lead to a biased result. As a step for robustness check, we repeat the above process using 10-fold cross-validation. In particular, for every iteration, we make use of 90\% of the dataset for training a propensity score model, matching, and measuring EATE. As the last step, we compute the 95\% confidence interval by averaging the 10 EATEs and discard the cases where the interval includes zero: this case indicates an effect's direction can be flipped for different folds. The reported EATE is the average of the 10 EATEs measured on the splits.

\subsection{Results}

Using the analysis framework, we investigate what effects are brought into user engagement on Twitter by editing tweets for news sharing. Note that the analysis framework is applied to each scenario that tests an effect of style $A$ (e.g., editing) compared to style $B$ (e.g., mirroring) for $K$ media outlet (e.g., NYTimes): we train a propensity model for each scenario separately, and we check semantic balance and robustness between its treatment group and corresponding matched control group. If a scenario cannot pass the balance check or the robustness check, we omit the case.

\subsubsection{Effects of Modifying Original Headlines}

First, we investigate whether mirroring a news headline is a good strategy for user engagement. The treatment group is  headline-tweet pairs where the tweet text is different from the news headline, and the control group is those which are identical to each other. Since user engagement distribution varies across media outlets as shown in Figure~\ref{fig:reaction}, we apply the propensity score matching to the headline-tweet pairs of each media separately.

\begin{figure}[t]
\centering
\includegraphics[width=0.95\columnwidth]{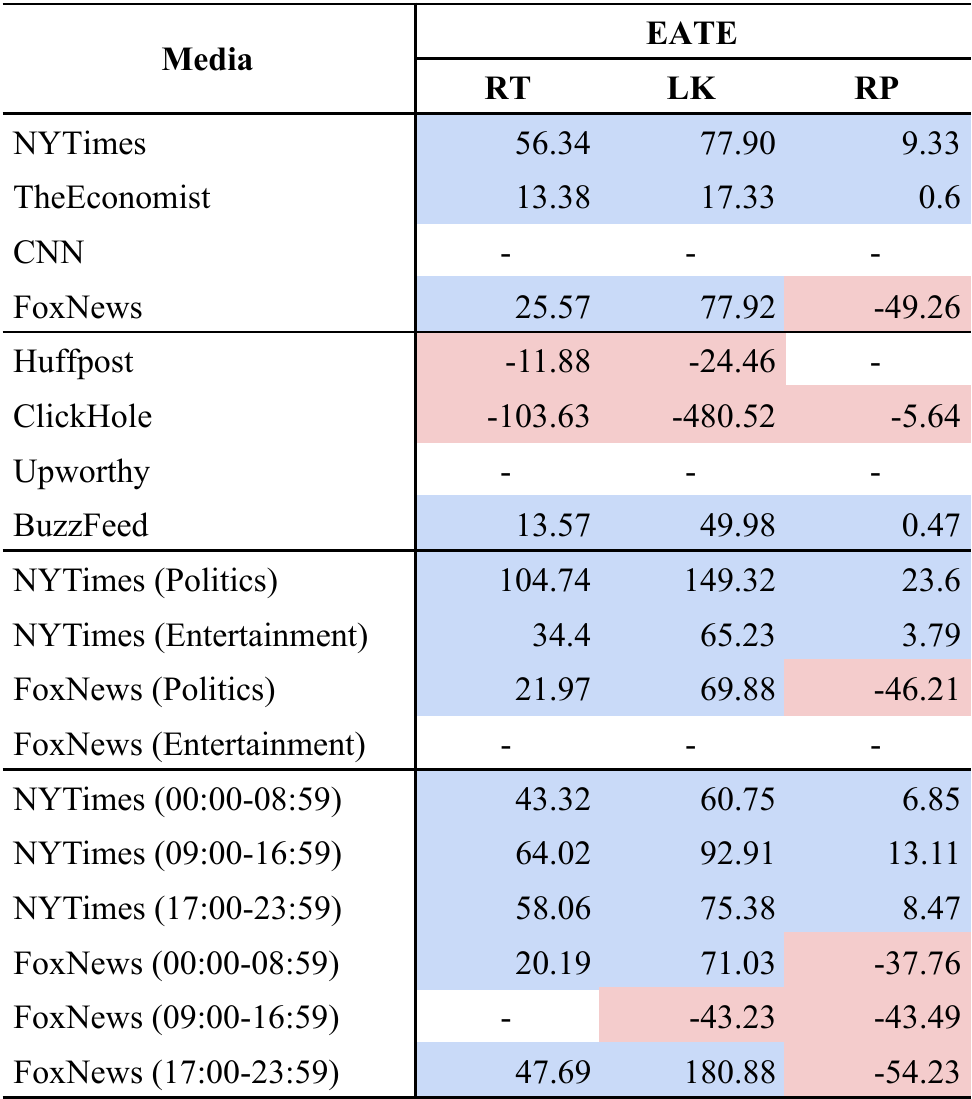}
\caption{Effects of editing tweets against news headlines on the amount of user engagement on Twitter. The blue-colored cell indicates a positive effect, and the red-colored one indicates a negative effect (RT: retweets, LK: likes, RP: replies).}
  \label{table:edit_effects}
\end{figure}

\begin{figure*}[ht]
\frenchspacing
\footnotesize
\centering
\includegraphics[width=\textwidth]{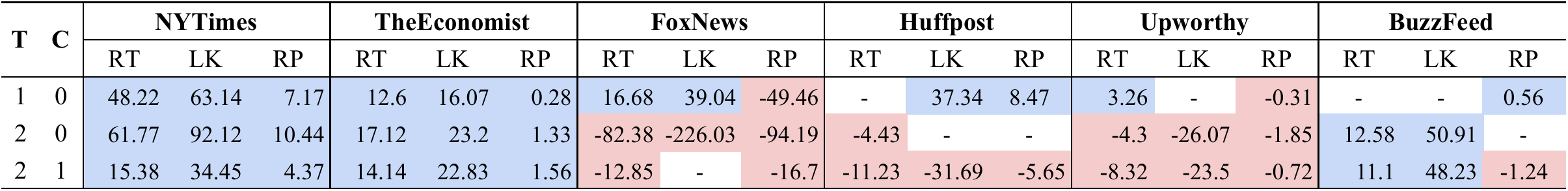}
\caption{Effects of the amount of content change in editing tweet messages against news headlines on user engagement on Twitter. Unsuccessfully matched entries are omitted (T: cluster index of treatment group, C: cluster index of control group, RT: retweets, LK: likes, RP: replies).}
\vspace{-2mm}
  \label{table:cluster_effects}
\end{figure*}

Figure~\ref{table:edit_effects} presents the EATE on the three variables of audience engagement, measured for the eight media outlets. According to balance check and robustness analysis, we exclude the CNN and Upworthy results. Here, we make three main observations. First, for the hybrid news media, changing news headlines is more likely to increase social media engagement than the mirroring style. For example, for NYTimes, the tweets edited from news headlines are on average more retweeted (+56.34) and liked (+77.90) than the tweets identical to news headlines. While the positive effect is similarly observed for TheEconomist, FoxNews exhibits a different pattern; the number of retweets and likes increased, but that of replies decreased. Second, for the online-only news outlets, editing news headlines for sharing tends to have diverse effects. We measure negative EATE values for Huffpost and ClickHole; that is, the mirroring strategy is more effective for the two media. Interestingly, Huffpost and ClickHole are media that changed news headlines the most and least (95\% and 19\%), respectively. By contrast, BuzzFeed enjoys positive effects like the hybrid media do. 
Third, the number of likes is always bigger than retweet counts, indicating that likes may be more likely to be influenced by editing headlines than retweets count. This pattern is consistent across all the media, and the finding is aligned with previous work on different levels of user engagement~\cite{aldous2019view}.

As discussed in the Related Works, the level of audience engagement could also be affected by other factors, such as topic of news and time of day. Thus, we further see if the estimated effects by editing tweets are generalizable against those confounding variables.

Considering the news section as a proxy of a broad topic, we first look into the effects of editing in politics (as hard news) and entertainment (as soft news) separately by repeating the whole analysis process for each scenario. For example, edited headline-tweet pairs in politics of NYTimes are only matched to the identical headline-tweet pairs in politics of NYTimes, according to the newly estimated propensity scores. 
Since NYTimes and FoxNews explicitly indicate the section information in their URLs\footnote{e.g., https://www.nytimes.com/.../\emph{politics}/article.html}, we focus on them in this experiment. The four rows in the middle of Figure~\ref{table:edit_effects} shows the EATEs measured on each section of those two media. For NYTimes, the direction of EATEs is the same across the politics and entertainment sections; that is, editing news headlines  is likely to increase user engagement, which is congruent with the observation from the whole NYTimes data. Similarly, for FoxNews, the directions of the EATEs measured on the politics section is the same as those from the whole. These observations suggest the generalizability of the effects by editing news headlines with controlling for the effects of topics.

As a second confounding variable, we consider the time of day a tweet posted. Following the practice of previous studies considering the effects of time on user engagement~\cite{orellana2016spreading}, we split the posting time into the three time blocks: 00:00-08:59, 09:00-16:59, and 17:00-23:59. We align the posting time with the Eastern Daylight Time (EDT) as the most of the global news media targets at EDT due to its importance in economy (e.g., U.S. stock market) and politics (e.g., Washington D.C.). For example, even though the headquarter of TheEconomist is located at London, they publish news articles following the EDT.

For headline-tweet pairs of NYTimes and FoxNews shared in each time block, we repeat the analysis process and present the results in the six rows at the bottom of Figure~\ref{table:edit_effects}. For NYTimes, the direction of EATEs is the same across the time blocks, which is also identical with that of EATE measured on the whole headline-tweet pairs of NYTimes. For FoxNews, the direction of EATEs is the same as that of the whole data for 00:00-08:59 and 17:00-23:59; yet, in 09:00-16:59, the effects on the number of likes are the opposite. We hypothesize the Twitter usage patterns in the working hours may affect the engagement patterns and future works could investigate how audience differ across time.

\begin{figure*}[ht]
\centering
\includegraphics[width=\textwidth]{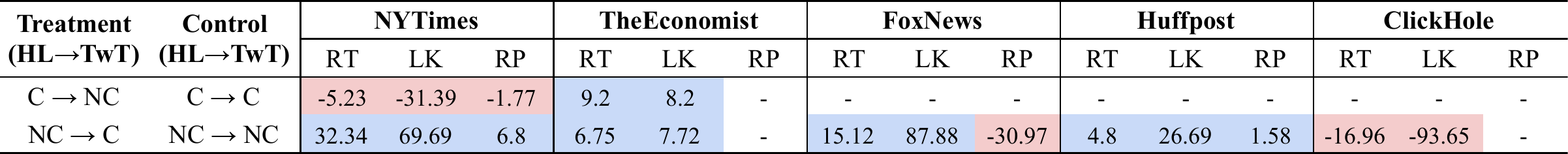}
\caption{Effects of controlling clickbaitness of news headlines (HL) into sharing tweets (TwT). Unsuccessfully matched entries are omitted (RT: retweets, LK: likes, RP: replies, C: Clickbait, NC: Non-clickbait).}
  \label{table:clickbait_effects}
\end{figure*}

Then, how much content should be changed (or kept) in terms of lexicons and semantics for effectively garnering user engagement on Twitter? To investigate whether an optimal style of content change exists, we apply the analysis framework based on the clusters identified in the earlier section (Figure~\ref{fig:cluster}), each of which could represent one of the editing styles: marginal change (Cluster 0), paraphrasing (Cluster 1), and semantic change (Cluster 2). For headline-tweet pairs in each media outlet, we consider the headline-tweet pairs of cluster $T$ as a treatment group and those of the other cluster $C$ as a control group. Again, a propensity model is trained on the dataset of each cluster pair of each outlet separately, and it is used for matching among the pairs published in a same outlet. Therefore, the varying popularity across media is automatically controlled. We run the experiments for all possible pairs of clusters but only report the cases where \textit{T}$>$\textit{C} due to the lack of space; we observe the direction of effects is always opposite when we swap the condition of treatment and control.

Figure~\ref{table:cluster_effects} presents the EATE of the matched results. Results display that there exists no single optimal cluster that leads to a positive EATE, and the trend varies across the media outlets. First, for NYTimes and TheEconomist, having Cluster 2 (Semantic change) as treatment group leads to a positive EATE. This suggests that, for the two media, editing news headlines by changing both words and semantics is an effective strategy to increase audience engagement on Twitter. 
In other words, both news media may well understand who are their audiences on Twitter and furthermore write highly engaging tweets that were often quite different from the original headlines. 
Second, Cluster 1 (Paraphrasing) leads to positive EATEs in Huffpost in comparison to the other clusters. Third, BuzzFeed tends to show the positive EATE for Cluster 2, and FoxNews tends to exhibit the positive EATE for Cluster 1; yet, the two media have the opposite effect for replies, suggesting that replies may have a different characteristics compared to the other two engagement measures.

In combination with the findings on the effects of the mirroring strategy in Figure~\ref{table:edit_effects}, the above results suggest that the optimal editing style varies across the news media outlets.

\subsubsection{Effects of Using Clickbait-Style Messages}

Next, we estimate the effects of clickbait for news sharing on Twitter. Note that we exclude identical headline-tweet pairs for the subsequent analysis to capture a distinct pattern from the impact of editing. Figure~\ref{table:clickbait_effects} shows the EATE on user engagement by sharing non-clickbait headlines with clickbait tweets and presenting non-clickbait news for sharing clickbait headlines. The clickbait label is annotated by the deep learning model described in \S 4.3. 

As shown in the second row in Figure~\ref{table:clickbait_effects}, sharing those with clickbait tweets (i.e., NC$\rightarrow$C) is likely to increase the number of retweets, likes, and replies for NYTimes, TheEconomist, and Huffpost, compared to sharing non-clickbait headlines on Twitter. This finding supports previous research's finding that clickbait can boost user engagement~\cite{blom2015click,park2020trust}. In FoxNews, the positive EATEs are observed for retweets and likes, but the EATE for replies is negative. The opposite trend of replies in FoxNews was repeatedly observed in the earlier analyses, including Figure~\ref{table:edit_effects} and \ref{table:cluster_effects}. This suggests FoxNews’s audience may react to shared tweets differently, and it calls for future investigations.

From the experiments on the effects of sharing non-clickbait tweets for clickbait news titles (i.e., C$\rightarrow$NC), we achieve successful matching across the three engagement measures only for NYTimes; three engagement measures have negative EATEs. Based on the results, we could assume NYTimes editors are not effective in sharing clickbait news articles with clickbait tweets, compared to the opposite case.

It is interesting to observe that TheEconomist shows the positive EATEs for both C$\rightarrow$NC and NC$\rightarrow$C. Given a news article, its social media manager may know a desirable style for being shared on social media. Since we do not have access to their internal guideline on presenting news on Twitter, we cannot explain the detailed underlying mechanism. Still, we can estimate the effectiveness by the causal inference framework.

We further evaluate whether the effects hold the same in the Politics and Entertainment sections for NYTimes and FoxNews for generalizability. In the analysis on the NYTimes Entertainment section, the EATEs are similar to those in the whole data, except for replies. On the contrary, in Politics, the effects become the opposite; sharing non-clickbait tweets for clickbait news in politics turns out to be beneficial for promoting engagement. The distinct direction of effects across the sections suggests there might exist desirable styles for different topics. In FoxNews, the section-level analysis also exhibits a different trend from that as a whole. In both sections, sharing clickbait tweets with non-clickbait news likely decreases the amount of user engagement. This contradicting observation suggests that FoxNews’ Twitter audience responds to clickbait tweets differently for Politics and Entertainment compared to news in other sections.

\section{Discussion and Conclusion}

Social media serve as places where people read and discuss news today~\cite{mitchell2018americans}. News organizations have run their media accounts to share their own articles on social media. 
Unlike traditional newspapers that readers can see headlines and body text at the same time, on social media, the main content is not shown to the readers, but a short text (e.g., tweet) should attract readers to click the link to read more~\cite{Park2020}. Therefore, it is crucial for news organizations to write an effective social media post to boost user engagement. 
The lack of available datasets and analysis frameworks, however, makes it challenging to evaluate which editing strategy is more effective in garnering user attention on social media in a systemic manner. 

As a first step to overcoming such limitations, we built a parallel corpus of news articles and tweets shared by the eight news outlets and examined how they edit news headlines for news sharing on Twitter (RQ1). 
The findings show that the media outlets employed diverse strategies in writing the social media messages. 
While mirroring a news headline to Twitter was a common strategy, the media outlets also made various levels of change on content; for example, the online-only media present clickbait tweets more frequently.

A natural follow-up question is which editing strategy effectively promotes user engagement for sharing news articles on Twitter (RQ2). To answer the question in a data-driven way, we utilized a systematic framework that incorporates deep learning with propensity score analysis; in particular, we used a deep learning model for predicting the likelihood of receiving a treatment condition, known as a propensity. The causal inference framework allows for estimating an editing style’s effect on audience engagement by matching counterfactual outcomes where the same article is shared with another editing style. The high performance of deep learning for text classification enables to mitigate the effects of covariates such as textual features more effectively in the matching process.

The findings of the RQ2 can be summarized as three:
First, editing a news headline was likely to increase audience engagement on Twitter than mirroring the headline in the four hybrid news media, which publish news articles through both offline and online channels. By contrast, in the news media that only keep online channels, the estimated effects of editing tweets were generally negative except for BuzzFeed. 
Second, there was no universal best strategy applicable to different media outlets in terms of lexical and semantic changes. For example, changing the original semantics of news headlines (Cluster 2) was estimated to be the best tactic for NYTimes, yet paraphrasing original headlines for sharing tweets (Cluster 1) was the best for Huffpost in terms of EATE.
Third, sharing tweets with clickbait-style messages was likely to increase audience engagement in the four outlets. This finding is congruent with a previous study showing rewriting news headlines with a clickbait text increased the amount of engagement in a Dutch news service~\cite{kuiken2017effective}. Yet, we observed the opposite direction of EATEs for ClickHole, which might suggest that the level of audience engagement is not just a function of editing styles but also dependent on who their audiences are. To test the hypothesis, future studies could characterize audience types of news outlets (e.g., socioeconomic status) and investigate how different editing styles are preferred by each group.

On top of the above observations from the eight media’s paired news-tweet dataset, we believe the overall analysis framework, from how to process the data to conduct propensity score analysis, could benefit any media outlets in practice to evaluate their internal guidelines. For example, in Figure~\ref{table:clickbait_effects}, TheEconomist has positive EATEs for both directions (i.e.,C$\rightarrow$NC, NC$\rightarrow$C), suggesting that they may know how to make use of clickbait effectively. In practice, news media outlets could evaluate their editing strategies by applying the analysis framework introduced in this paper. We, therefore, release our systematic analysis framework as an easy-to-use toolkit.

\subsubsection{Limitation and Future Direction}

Although we consider diverse news media from hybrid to online-only and left to right in this study, additional studies with more and diverse news media are essential for evaluating the observations’ generalizability. We hope the shared analysis toolkit can serve as a starting point in the following studies. Another weakness of this study is an inherent limitation of the propensity score analysis, which is the risk of unobserved covariates. Based on the findings of the literature on news engagement, we tried to minimize the risk through various comparisons and robustness analyses. Last but not least, an editing style with a positive EATE suggests adopting the style may boost user engagement, but simultaneously it could have an adverse effect; future studies could examine the long-term impact.

Beyond the news domain, future studies could extend our analysis framework to other cross-platform sharing activities~\cite{park2016persistent,kholoud2019}. For example, how could researchers share their research papers on social media to effectively draw attention and achieve more citations in the long run? Mirroring the paper title may not be the best strategy because a scientific paper is usually written in a formal language. Our framework can be used to quantify which text styles would be more effective. Another exciting research direction is to automatically generate a social media post when a news article is given. Training a naive sequence-to-sequence model might not work well as there exist diverse headline-to-tweet mappings, as shown in this study. Future researchers could develop controlled generation technologies such as \citet{hu2017toward} for handling such diversity in the mappings.

\section{Acknowledgments}
This work was started when KP, HK, and JA worked in QCRI. This work was partially supported by the Singapore Ministry of Education (MOE) Academic Research Fund (AcRF) Tier 1 grant.

{
\bibliography{main}
}

\end{document}